# Gait recognition via deep learning of the center-of-pressure trajectory

Philippe Terrier[1,2,]

Affiliations:

[1] Haute-Ecole Arc Santé, HES-SO University of Applied Sciences and Arts Western Switzerland, Neuchâtel, Switzerland

[2] Department of Thoracic and Endocrine Surgery, University Hospitals of Geneva, Geneva, Switzerland

ORCID Number:

0000-0002-3693-4505

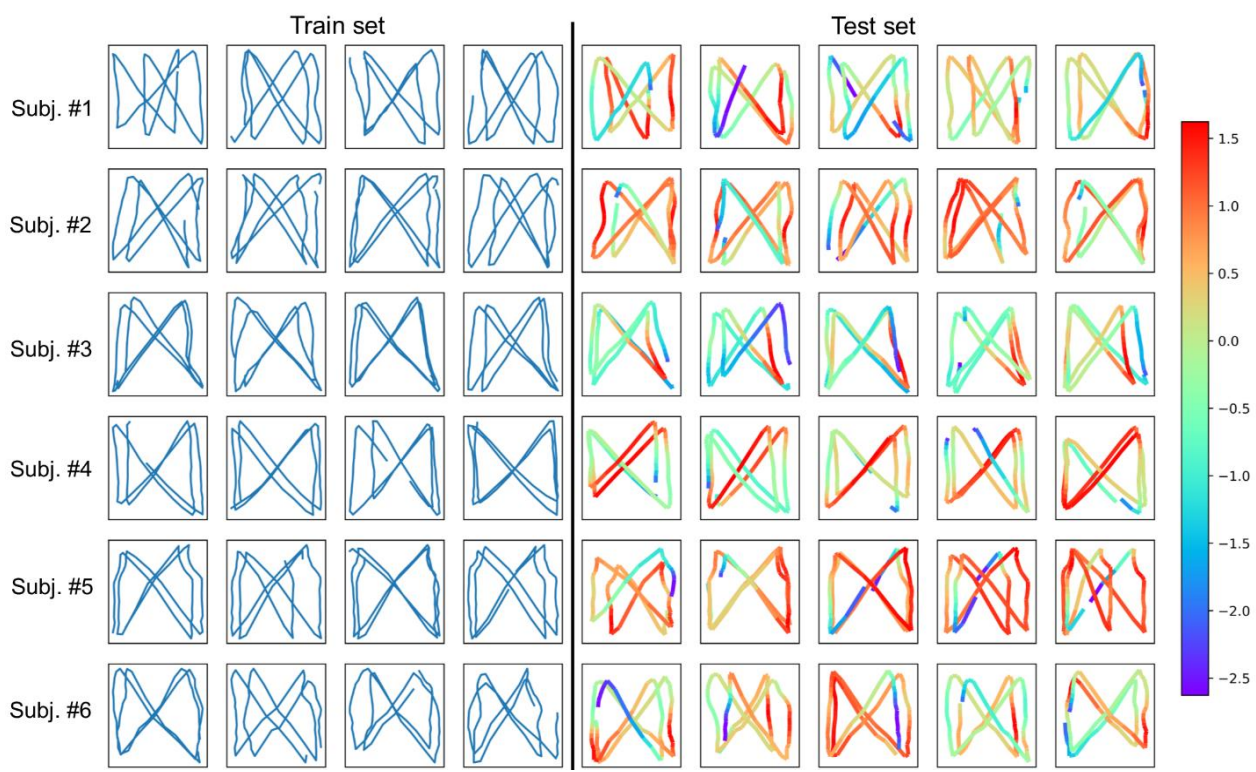

**Fig. 6.** *CAM analysis. Twenty-four segments were randomly selected from the training set containing the gait data of six participants (left columns). These segments were used to fine-tune the last layers of the convolutional neural network (CNN) that was pre-trained on the training set of the 30 other participants. This CNN classified 30 segments drawn randomly from the test set with 100% accuracy (right columns). Class activation mapping (CAM) was performed on each sample of the test set. Color coding shows which parts of the signals are prioritized to be used by the CNN to perform classification. Warm color (red, orange): high focus; cold colors (green, blue): low focus.*

**Abstract**

**The fact that every human has a distinctive walking style has prompted a proposal to use gait recognition as an identification criterion. Using end-to-end learning, I investigated whether the center-of-pressure trajectory is sufficiently unique to identify a person with a high certainty. Thirty-six adults walked on a treadmill equipped with a force platform that recorded the positions of the center of pressure. The raw two-dimensional signals were sliced into segments of two gait cycles. A set of 20,250 segments from 30 subjects was used to configure and train convolutional neural networks (CNNs). The best CNN classified a separate set containing 2,250 segments with 99.9% overall accuracy. A second set of 4,500 segments from the six remaining subjects was then used for transfer learning. Several small subsamples of this set were selected randomly and used for fine tuning. Training with two segments per subject was sufficient to achieve 100% accuracy. The results suggest that every person produces a unique trajectory of underfoot pressures and that CNNs can learn the distinctive features of these trajectories. Using transfer learning, a few strides could be sufficient to learn and identify new gaits.**

# 1    Introduction

Human beings move through their environment using repetitive movements of the lower limbs, such as walking or running. The sequence of these movements constitutes one's gait. One gait cycle (or stride) is created by the alternation of stance and swing phases performed by the legs. The gait pattern is constrained by biomechanical and energetic factors [1]. Furthermore, each individual has a unique gait signature that can be used for identification purposes, in a process known as gait recognition [2]. The most interesting property of gait recognition is that it can identify subjects without their knowledge or approval, contrary to other biometric methods such as fingerprint or iris recognition.

Video-based methods dominate the field of gait recognition [3, 4]. Researchers can benefit from numerous easily available video databases [5], of which the largest contains more than 10'000 individuals [6]. Recent advances show a recognition rate between 90% and 95% in optimal viewing conditions; but the accuracy drops under challenging conditions (such as occlusions, view variations, or appearance changes) [4]. Wearable inertial sensors have also been proposed for recognizing gait [7, 8]. These sensors are used extensively in biomedical applications for gait analysis, and, hence, benefit research efforts substantially [9]. High recognition rates (>95%) have been observed under laboratory conditions [10], but a lower accuracy (~70%) has been reported in more realistic datasets [11].

Analyzing the force that a walking individual applies to the ground has also been proposed for identifying people, an approach referred to as footstep recognition [12]. Different gait features can be extracted through force sensors embedded in the floor: the temporal sequence of the ground reaction force (GRF) [13–15], the shape of the foot on the ground (footprint) [16], or the trajectory of the center of pressure (COP) [17]. The COP is the point at which the resulting vertical force is applied (i.e., the integrated vectorial pressure field). Promising results have been obtained, with classification rates higher than 95% [15, 16]  (see also Section 2); however, the number of footstep recognition studies are still low, especially those including COP analysis. In particular, COP trajectory has never been used alone for identification and verification aims. In addition, most of the footstep recognition studies included only a few individuals performing a limited number of strides [2, 12].

Recognizing people through their gaits relies on the analysis of multiple complex features. Neural networks are therefore helpful for this task [18]. Convolutional neural networks (CNNs) are used with great success for video-based gait recognition [19, 20]. CNNs are especially well suited for working with images as a result of their strong spatial dependencies in local regions and a substantial degree of translation invariance. Similarly, time series can exhibit locally correlated points that are invariant in translation. The performance of deep CNNs for the classification of uni-dimensional or multidimensional time series is attested [21, 22]. Like for image classification, CNNs can extract deep features from a signal's internal structure. CNNs are potent tools for bypassing feature engineering in signal processing tasks (end-to-end learning) [22]. However, CNNs, as other artificial neural networks, require hundreds of examples in each class for an efficient learning; therefore, they have not been applied in footstep recognition studies so far, owing the difficulty to collect many strides with sensing floors or force platforms.

The idea behind this study was to harness the COP trajectory on the ground to identify individuals. To this end, the feature extraction capabilities of CNNs were used to classify gaits. I applied state-of-the-art CNNs for supervised end-to-end learning. An instrumented treadmill was used to collect hundreds of consecutive strides. The objective was to provide a proof-of-concept for the notion that measuring the COP alone can be used for biometric purposes. First, I assessed the classification accuracy of the method based on many strides (> 500 per participant) in both identification and verification scenarios. Second, I used transfer learning to explore whether new gaits could be successfully classified when learned from only a few strides.

## 2    Related works

In 2004, Jung et al. [17] suggested combining static and dynamic foot-pressure features to identify walking individuals. They used a mat containing $40 \times 80$ pressure sensors ($1 \times 1$ cm$^2$ resolution) to record footprints and COP trajectories and collected one-step footprints from 11 participants. Forty footprints were recorded from each subject for two months. An overall classification accuracy of 98.6% was obtained through the use of hidden Markov model and Levenberg-Marquart learning methods. One strength of the study was that data were recorded over a long period of time and tended to demonstrate that foot-pressure features are time invariant to a substantial degree. Study limitations included the small sample size and the fact that the subjects had to walk barefoot, making the method difficult to apply in practice.

In an article published in 2007, Suutala & Röning described a method for using a pressure-sensitive floor to identify individuals [18]. They covered the floor of their research laboratory with 100 m$^2$ of an electro-mechanical film sensitive to pressure and collected gait pressure data from 10 individuals. The researchers focused on the pressure intensity profiles, rather than on pressure trajectories or foot shape, applying different classifying algorithms, of which support vector machines and multilayer perceptrons were the most accurate. The most outstanding finding was that 95% classification accuracy was achieved when multiple features collected from several consecutive steps were combined. Despite the small sample, this study demonstrated the technical feasibility of using an instrumented floor for biometric applications.

The purpose of Pataky et al. in their 2012 study [16] was to analyze dynamic foot pressure patterns through plantar pressure imaging using a pedography platform that recorded footprints with 5mm resolution. They recruited 104 individuals and collected 1,040 barefoot steps. Several pre-features characterizing the pressure patterns where extracted, and dimensionality reduction technique was applied. A one-fold nearest-neighbor classifier was applied with cross-validation. The results show that the best feature was the pressure-time integral (PTI), with a classification rate of 99.6%. Overall, this study demonstrated that plantar pressure patterns are highly unique among individuals. The major practical limitation was that they investigated unshod walking.

In 2015, Connor studied foot pressure data collected from 92 subjects walking either unshod or shod on a pressure mat (255 x 64 pressure sensors) [23]. The purpose was to evaluate the classification and identification performance under three scenarios: 1) a "barefoot" scenario, in which the recognition system classifies only barefoot data; 2) a "same shoe" scenario, in which the system classifies footsteps of subjects wearing the same footwear throughout; and 3) a "different shoe" scenario, in which the system learns from examples of individuals walking with one type of footwear, and then evaluates footsteps when they walked with a different type of footwear. Connor then assessed many pressure-derived features and combinations thereof, including COP trajectories, GRF timing, footprint images, and general gait features. The results revealed that for the most difficult scenario (i.e., #3), it was possible to achieve a classification accuracy of 90.5%. Interestingly, COP parameters were among the most important features for an optimal classification in both shod scenarios (#2 and #3). In addition, gait metrics such as cadence, step length, toe-out angle, and foot length also played a substantial role. In short, this study demonstrated that COP trajectory might be appropriate for classifying shod walking gaits.

## 3    Methods

### 3.1 Data collection and pre-processing

The data used in the present study were collected by the author in a previous study aimed at analyzing the influence of synchronization to external sensory cues on gait variability [24, 25]. Thirty-six healthy adults participated in the study: 14 men, 22 women; means and standard deviations of their individual characteristics were: age 33 years (10), body height 1.72 m (0.08), and body mass 66 kg (13).The experiment consisted of 30 min treadmill walking under three cueing conditions: no cueing, auditory cueing, and visual cueing. The participants wore their customary shoes, but high heels were forbidden. The treadmill was instrumented with a force platform consisting of six load cells that retrieved the intensity and the position of the vertical force exerted by the subject walking on the treadmill surface [26, 27].

On a treadmill, the COP trajectory of a walking individual has a typical butterfly-like shape [28], as shown in Fig. 1. The "wings" of the butterfly correspond to the stance on a single foot, whereas the central crossing corresponds to the double-support phase when the body weight passes from one foot to the other. For a dynamical representation of the process, refer to a short video published in the supplementary material of a previous article [29].

The 500 Hz two-dimensional (2D) positional signals were low-pass filtered at 30 Hz and down sampled to 50 Hz. Each stride was identified in the raw signals [24, 26]. Five-hundred strides were kept for each cueing condition and each participant. These 500-stride time series were resampled to a uniform length of 20,000 samples, that is, 40 samples per gait cycle. The aim was to standardize the average stride duration among participants. Hence, the dataset contains three 2D signals of 20,000 sample length for each of the 36 participants, for a total of 54,000 gait cycles.

Each 2D signal of 20,000 samples was split into three parts: the first 16,000 samples were added to the training set, the next 2,000 samples to the development set (dev set), and the last 2,000 samples to the test set. The signals were stacked across subjects and conditions in arrays of three columns corresponding to the x-axis, y-axis and subject's identification (ID) number (1–36).

Finally, a non-overlapping sliding window algorithm sliced the arrays into small segments of 80 samples each (i.e., two gait cycles, or four steps). Four random examples of these segments are shown in Fig. 2. Each segment was labeled with the subject's ID, which was converted into categorical format. Two ensembles of sets were created, the first for developing and testing the CNN models, containing the data of 30 subjects (18,000 segments in the training set, and 2,250 segments in the dev and test sets), and the second for transfer learning, containing the data of six subjects (4,050 segments in the training set, and 450 segments in test set).

### 3.2 Software and Data Availability

GPU computing, with Tensorflow and Keras on Python 3.6, was used for CNN development. Other Python libraries were Pandas, Numpy, Hyperopt, Hyperas, Matplotlib, and Scikit-learn. Some preliminary steps (raw signal filtering) were conducted using MATLAB. The raw data are available on FigShare [30].

### 3.3 CNN

The overall network architecture is shown in Fig. 3. I designed a CNN based on stacked one-dimensional (1D) convolutional layers systematically followed by batch normalization [31] (not shown in Fig. 3 for the sake of simplicity). Zero padding was used to ensure identical input/output sizes among layers. Maximum pooling layers were used for reducing dimension (Fig. 3). I used a CNN architecture that included skip connections, i.e., a ResNet-like architecture [32] that has also been advocated for time series classification [22]. These shortcut paths between non-consecutive layers allow a better flow of information in a deep CNN, preventing in part the issue of vanishing/exploding gradients [33]. When shortcuts required dimension adjustments, one-fold convolution layers were applied with the appropriate number of filters (depth adjustment) or adjusted stride (temporal reduction). Nonlinearities were introduced through the use of activation layers interleaved as recommended for the ResNet architecture (Fig. 3).

In addition to the standard 1D convolution layers, I tested whether depthwise separable 1D convolution layers could provide a valuable alternative (Xception architecture [34]), hereafter referred to as sepCNN. These layers combine pointwise and depthwise convolutions, resulting in a factorized version of standard convolutions. SepCNNs use fewer parameters and are computationally more efficient [35], possibly a major advantage for practical applications. I used a similar architecture for both CNNs and sepCNNs, but their hyperparameters were tuned independently (see Section 2.4).

The loss function was categorical cross-entropy. The Nadam algorithm was chosen as the mini-batch gradient descent optimizer [36]. Nadam is a variant of the classical Adam algorithm [37], but with Nesterov momentum [38]. The recommended parametrization was used. The models were fitted using a mini-batch size of 256. The metric was overall accuracy (correct classification rate), i.e., the number of segments assigned to the correct person over the total number of segments.

### 3.4 Hyperparameter tuning and model testing

Table 1 summarizes how the CNN hyperparameters were tuned. Regarding the model architecture, both the number of filters and the number of intermediate blocks (Fig. 3) were adjusted. I also evaluated two different approaches for the final layers: the first using a classical combination of dense–dropout–softmax layers, and the second using a global average pooling layer [39] preceding the softmax layer.

Regarding activation, four algorithms were tested: ReLU [40], LeakyReLU [40], PReLU [41], and trainable Swish [42]. Swish is a recent algorithm similar to the sigmoid-weighted linear unit proposed in [43], but with a trainable parameter. Regarding convolutional layer initialization, two algorithms were tested, the so-called Glorot-normal (a.k.a., Xavier-normal, [44]) and the He-normal [41]. Regarding L2 regularization, the optimal weight decay ($\lambda$) was searched for between $10^{-7}$ and $10^{-3}$. Finally, the optimal initial learning rate was searched for between 0.1 and 2 times the recommended value of 0.02, i.e., between 0.002 and 0.04.

Bayesian optimization was used to search through the hyperparameter space. More precisely, I applied the tree of Parzen estimators (TPE) algorithm [45] as implemented in the Hyperopt library [46] and its Keras wrapper, Hyperas. [42] The overall accuracy on the dev set was used as the metric. Three-hundred trials were run and the combination that provided the highest accuracy was chosen.

A final training set was built by concatenating the train and dev sets. The accuracy of the best setting (Table 1) was assessed on the test set. An early-stopping algorithm was used to reduce the training time. The assessment was repeated 10 times to take model stochasticity into account (Table 2).

### 3.5 Transfer learning

The objective was to analyze the ability of the models to generalize on new gaits. The aim was to find the minimal number of strides required to tune a pretrained model for a correct classification of gaits of previously unseen individuals. Indeed, for future efficient applications, it is important to know whether gaits can be learned based on only a few strides.

The gait data from the six subjects not included in the model development were used. I applied the principles of transfer learning: the best models trained on the 30 subjects were fine-tuned on the basis of the new gait data. First, most model parameters were frozen; that is, their trainable attributes were set to *false*; only the weights of the last two convolutional layers, as well as the parameters of the batch normalization layers, were kept trainable (see Fig. 3). Then, the output (softmax) layer was replaced with a new layer with six neurons to match the new classification task. Regarding optimization, the learning rate was set to 0.0002 for the fine tuning of the weights of the last layers.

To investigate the discriminative power of the new model, the output of its untrainable part, as well as the output of the two trainable layers, were analyzed using t-distributed stochastic neighbor embedding (t-sne, Scikit-Learn implementation) [48]. T-sne is a nonlinear dimensionality reduction algorithm that embeds high-dimensional data in a low-dimensional space of two dimensions for visualization purpose. Here, t-sne helped visualize whether the segments of each individual were clustered together. New CNN and speCNN models were trained with the 4,050 segments of the training set (10 epochs, batch size = 32) and then inferred on the 450 segments of the test sets. T-sne reduced the dimension of the flattened layer outputs to 2D (Fig. 4).

The new models were then trained and tested on very small samples as follows: from the full test set containing 450 segments, 60 of them were randomly selected (10 per subjects); then, a variable number of random segments were drawn from the training set of 4,050 segments: six, 12, 30, or 60, that is, one, two, five, or 10 per subject. Finally, the overall accuracy on the test set was computed. Fifty repetitions of this procedure were conducted for each segment number (for a total of 200 repetitions). Boxplots were used to show result distributions (Fig. 5).

### 3.6 Verification

In most biometric applications, the goal is not necessarily to identify an individual, but rather to verify whether an individual is an authorized user or an impostor. To test this type of verification scenarios, 100 different training & test sets were built for challenging the best CNN in impostor rejection tasks. First, a full training set including 97,000 steps (24,300 segments) of the 36 participants was gathered. A test set including 11,000 steps (2,750 segments) was built as well for testing the performance of the classifier. The segment labels identifying each participant (ID 1-36) were modified as follows: each participant could be considered as an authorized user (label = 1), or as an impostor (label = 0); the repartition between authorized users and impostors was randomly chosen. Four different levels of repartition were chosen: 10 authorized users vs. 26 impostors, 15 vs. 21, 20 vs. 16, and 25 vs. 11. Twenty-five repetitions for each repartition level was performed, each repetition including a random assignment of individuals between groups. As for transfer learning, the best pre-trained CNN was modified for the new classification task. CNN weights were frozen, except those of the two last convolutional layers. Given the binary nature of the classification task, the output layer was replaced with a logistic classifier (sigmoid activation). The mini-batch size was 128. The learning rate was 0.0002. Ten epochs were performed to fine tune the pretrained CNN. Two different performance indexes adapted to unbalanced classes were used: the area under the receiving operator characteristic curve (AUC), and the equal error rate (EER) [49].

### 3.7 Class activation mapping

Class activation mapping (CAM) was used to develop a better understanding of how the CNN classified the gaits [50]. First developed for computer vision, CAM indicates the discriminative regions of an image that are used to identify the class. Time series can also be analyzed using this method [22, 51]. In this case, CAM shows the time interval of the signal that was used for classification. CAM takes advantage of the global average pooling layer occurring after the last convolutional layer. Given this simple connectivity structure between the softmax layer and the outputs of the last convolutional layer, softmax weights can be back-projected onto the feature maps. Indeed, the softmax weights corresponding to one class indicate the relative importance of each feature for that class. Weighted feature maps are then summed to provide the final CAM, which can be up sampled to the original input size for an optimal interpretation.

I modified the best pretrained CNN for CAM analysis. Indeed, the temporal resolution (six points) was too low owing to successive max-pooling layers. The last max-pooling layer was removed, and the last convolutional layers were replaced. The temporal resolution was therefore 20 points. Raw CAMs were standardized and up sampled to the original input size (80) using spline interpolation. A small dataset of 24 segments from six participants was chosen randomly as the training set. The new model was then fitted (25 epochs, batch size of six). CAM was computed on 30 segments selected randomly from the test set (Fig. 6).

## 4 Results

The last columns of Table 1 display the optimal combination of hyperparameters for both CNNs and sepCNNs. The optimal CNN model consisted of 12 1D convolutional layers and the optimal sepCNN model of 9 separable 1D convolutional layers. The number of parameters were 0.7 and 2.1 M for the CNN and the sepCNN respectively. Detailed drawings are available in the online supplementary material.

Table 2 shows the accuracy on the test set for 10 different trainings with 20,250 segments. For 2,250 segments in the test set, the CNN misclassified 3.5 segments and the sepCNN, 2.5 (medians). On average, model training required 28.1 epochs for CNN and 20.3 epochs for sepCNN.

With the pre-trained models adapted for classifying new gaits from the six remaining individuals (transfer learning), the accuracy was 100% for both CNN and sepCNN when the full sets were used. The t-sne analysis (Fig. 4) shows that while the untrainable portion of the models could not, as expected, separate features (-3), the last two convolutional layers (-2 and -1) were able to separate individuals fully.

The overall results of transfer learning on small sub-samples are summarized in Fig. 5. Using only one segment per subject for training both CNN and sepCNN was sufficient to achieve 100% accuracy (median over 50 repetitions), that is, a correct classification of 60 segments over 60 in the test set. Adding more segments reduced the number of low accuracy outliers. Overall, CNN outperformed sepCNN. CNN reached 100% accuracy in 182 of 200 trials (91%), whereas sepCNN achieved 100% only in 172 of 200 (72%).

The results of the verification experiments are shown in Table 3. Among the 100 experiments, the AUC values were systematically near one, which demonstrates the high capability of the CNN to separate between authorized users and impostors. EER values lied between 0.25% and 0.33%, (average: 0.29%), which also attests that the CNN can deal with verification scenarios.

Fig. 6 shows the CAM results, which allows us to visualize which parts of the COP signals contributed the most to the classification. The left columns show the segments included in the training set, and the right columns show the segments of the test set that the adapted CNN succeeded in classifying with 100% accuracy. Each row contains the data of one subject. Warm colors show which part of the COP signal was used by the CNN to perform the correct classification. In one case (subject #3), the CNN focused on a prominent pattern of the lower right part of the trace, which corresponds to the terminal stance phase. In another case (subject #4), the focus was on the diagonal, which corresponds to the double-support phase (or pre-swing). The other cases exhibit inconsistencies among samples, making the interpretation difficult.

## 5 Discussion

The aim of this study was to highlight the potential of COP analysis for biometric purposes. I investigated whether the COP trajectory could discriminate among 36 individuals. The learning of 675 gait cycles per participant under supervision using deep CNNs enabled the classification of 75 previously unseen strides with an overall accuracy of 99.9%. In verification scenarios, the best CNN was able to detect impostor gaits with an EER of 0.29%. Transfer learning results showed that pre-trained CNNs can learn new gaits successfully when fed with only two to four strides.

With only three to four misclassified items over 2,250 attempts, CNNs were found to perform very well in the task of recognizing individuals through their COP trace on the treadmill belt (Table 2). As well, in verification scenarios, impostors can be identified with a high accuracy (EER 0.29%, Table 3). Table 4 compares these results with those of previous representative footstep recognition studies. Classification of COP traces via CNN seems to perform equally or better than other approaches. Note that previous studies relied on complex procedures of feature engineering and data reduction before classification. In contrast, here, CNNs were directly fed with raw COP signals. It is also worth noting that the previous studies that reported the highest accuracy [16, 23] analyzed unshod walking, and have, hence, a limited applicability in real-life situations. The outside-the-lab application of the 3D GRF method [14, 15] is also questionable, given that it uses complex and expensive force platforms that can cover only a very limited surface.

The high performance of CNNs was obtained with only a very slight regularization, with no dropout. Interestingly, the novel Swich algorithm revealed itself as the best activation method, in line with the results obtained for image classification [42]. The hypothesis that the use of depthwise separable convolution could favor a simpler and more computationally efficient model was not verified. Indeed, although slightly more accurate, the best sepCNN had over 2 M parameters versus 700 k for the best standard CNN. SepCNN appears to require far more filters per layer than CNN, which outweighs the fact that separable convolution requires fewer parameters. However, further analyses are needed to improve model architecture and select appropriate hyperparameters more adapted for sepCNN.

Traditionally, CNNs used for image classification, such as AlexNet [52] or VGG [53], consist of two distinct parts: a feature extraction part (convolutional layers) and a classifier (fully connected, or dense, layers). In transfer learning, the feature extractor is frozen, and the dense classifier is replaced and trained for the new task. Modern CNN implementations, since GoogLeNet [54], take advantage of fully convolutional networks, in which a global average pooling layer summarizes the feature maps before the output layer. This approach can also be applied to time series classification [22]. Hyperparameter tuning results (Table 1) highlighted that the second solution was better for the intended task of gait classification. The t-sne analysis (Fig. 4) confirmed that the last two convolutional layers could classify new gaits on their own, without the help of dense layers. This is a major advantage in terms of computational efficiency, because convolutional layers require far fewer parameters than fully connected ones.

The high accuracy obtained in classifying gaits was likely due to the similarity of distribution between the training and testing sets. Under laboratory treadmill conditions, consecutive gait cycles are expected to have a high resemblance among them. However, an extended time stability of gait patterns is supported by several studies [17, 55–58]. In 2018, Nuesch et al. [56] showed that foot rotation and step width had intraclass correlation coefficients (ICC) greater than 0.93 when two measurements on different days were compared. These recent results accord with older findings [58]. Another recent study [57] also showed that most gait parameters, including GRF, are reproducible from one day to another (ICC > 0.95). Hence, the intrinsic variability of foot pressures appears to be compatible with biometric applications.

Which part of COP's typical shape is important for recognizing individuals? Answering this question would be helpful in understanding which phase of the gait cycle is the most discriminative. In an attempt to answer this question, I applied the CAM technique (Fig. 6). Overall, it seems that no gait phase stands out as a privileged CNN target. When a pattern is sufficiently unique, as for subjects #3 and #4, the CNN uses it for classification, but in other cases (#2 and #5), the CNN appears to prefer a more global approach. This illustrates the great adaptability of the CNNs, which are able to extract the most useful features for a task without preliminary feature engineering.

One prominent particularity of the present study was the use of an instrumented treadmill to collect gait data. The main advantage was that a large number of steps (108,000) were recorded in a constant environment. It was thus possible to use deep neural networks, which are known to require substantial data to fulfill their full potential [59]. The drawback was that treadmill gaits can differ from standard (overground) gaits [60]. Although this difference is deemed to be small [61], future investigations should focus on overground walking. It is worth noting that the butterfly-like diagram of COP can also be obtained in overground walking by subtracting the average speed vector from the trajectory [62].

Because the continuous gait data were segmented in small segments of two strides (Fig. 2), a pressure-sensitive floor of 3 m length could capture enough foot pressure data to identify individuals: with such a length, indeed, the recording of two consecutive strides is possible even for fast walking [63]. Transfer learning results also highlighted that two strides could be sufficient to reconfigure a pretrained CNN to classify previously unseen gaits. Indeed, when feeding the reference CNN with only one segment collected from the six participants not used in the CNN design phase, the classification accuracy of 60 segments of the test set reached 100% in most repetitions (Fig. 4). The first layers of the CNN very likely learned to separate the general features of the butterfly-like shapes that are common to everyone. Based on this preliminary feature separation, the last layers can easily recognize new gaits by learning details that are typical of each individual. Transfer learning is therefore a potent tool that extents the use of deep CNNs to datasets of any sizes.

The gait dataset used in the present study was reemployed from a study aiming at a better understanding of gait variability under different conditions that modified attentional demand [24]. Attention changes are frequent in free-living walking: for example, a higher attention to gait is required when navigating through crowded environments. The results of transfer learning strongly suggest that these attentional changes did not impacted gait recognition. Indeed, when only one segment per individual is used to learn new gaits, it must come from only one experimental condition: it was however possible to correctly classify gaits from other experimental conditions. In other words, the COP trajectory seems to remain constant even if the degree of attention dedicated to gait control is modified.

Two techniques are in use for measuring the COP of walking individuals: 1) reaction force measured through strain gauges (force platform), or 2) pressure field measured through a grid of pressure sensors. In this study, I applied the first solution. Instrumented walkways including force platforms for analyzing overground walking exist for medical applications [64]. The COP trajectory is computed by aggregating the signals of strain gauges that are placed on the edges of the walkway every 1 to 2 m. COP trajectory assessment requires the measure of the vertical force only (i.e., a 3-component force platform [65]), which is a major simplification as compared to methods relying on 3D GRF signals that require a 6-component force platform [65]. The major issue is that the correct COP position can be obtained only when one individual at a time steps onto the sensitive area. The second solution exists for both treadmill [29] and overground walking [62] and has been privileged for biometric applications, given that footprint shape is also helpful for recognizing gaits (see Section 2). In this case, COP is retrieved using the weighted average of the pressure sensor outputs. The gait of several persons can be simultaneously recorded on the same sensitive area, provided a pre-processing algorithm separates individual footprints [66]. The drawback of the pressure sensor grid is a complicated technical setup that generates a large quantity of raw data. Indeed, hundreds to thousands of sensors are required to cover a large area, each sensor generating its own pressure signal.

As other footstep recognition studies (Table 4), the present study was conducted under controlled laboratory conditions. This is not fully representative of spontaneous walking in habitual environments. One major issue is that the participants walked at a constant preferred speed. Although preferred walking speed is known to be constant (3%–4% variation [55]), voluntary control and changing conditions (such as slope or crowding) can modify it. The impact of these speed changes on the COP trajectory requires further investigations. If any, the gait gallery used as reference for future identifications could include gaits collected at different speeds. A second major issue concerns the footwear. As shown by Connor [23], identifying people wearing different shoes in the gallery and in the probe sets is a challenging task. Clarifying footwear effects on the COP trajectory is therefore a priority for future studies.

How well the COP method could perform in comparison with the reference methodology of gait recognition, namely video-based gait recognition? Both methods share the interesting ability to identify or verify users without their knowledge or cooperation. The performance of both methods can be potentially affected by intrinsic variability of gait patterns induced by long term changes (age, gait disorders, weight gain) or transient modifications (walking speed, carrying conditions, injuries, clothing and footwear) [4]. On the contrary, the COP method is not affected by extrinsic sources of error known to affect video-based recognition, such as changes in lightning conditions, viewing angle, and occlusions. Furthermore, regarding storage and computational needs, COP recording has the advantage of generating far fewer data than videos and hence could be more efficient in real-time applications.

That said, video-based gait recognition brings the overwhelming advantage of exploiting the data of countless video surveillance networks installed all around the globe. On the contrary, COP analysis cannot exploit already installed infrastructures and could be used in specific situations where video monitoring is difficult or unwanted. For instance, in many countries, video monitoring in the workplace is restricted for protecting employees' privacy.

Short-term gait tracking is a potential application for gait recognition based on COP trajectory, for example in the context of high-security buildings. Let us imagine the following scenario. A high-tech company has a research center in which future high-profit products are developed. The company wants to protect its scientific assets from industrial espionage with minimal constraints for the scientists. For the sake of demonstration, let us also imagine that scientists are reluctant to be filmed via a camera network because they feel that continuous spying harms their freedom of research. At the building entrance, an individual is first identified (through ID card, face recognition, or other means), and a reference sample of his or her gait is collected on an instrumented walkway. The recognition system thereby acquires a gallery of gaits of all individuals currently authorized to be in the building. Strategic areas of halls and hallways are equipped with pressure-sensitive floors that continuously track gaits of people walking by. Specific doors can be opened on-the-fly for authorized personal without user interaction. Access to high-security rooms can be audited and potential intruders rapidly localized. Privacy is respected, because only employees' localization when they are walking at some specific places is monitored, and not what they are doing. In this scenario, the footwear inconstancy issue is bypassed, because it can be expected that individuals keep the same shoes on throughout the day. The technology for this biometric approach is already available, but a cost-benefit analysis must be further conducted.

## 6 Conclusion

Studies that have used foot pressure data to identify individuals have relied primarily on footprint shapes or GRF [18, 23]. The principal finding of the present study is that COP trajectory alone can very likely identify people with a high confidence. The method requires simpler—less expensive—force platforms that those used for 3D GRF recognition [14, 15]. Alternatively, COP measurement could also be achieved by means of a pressure sensor grid [23]. The number of consecutive strides required for both constituting a reference gallery (four strides with transfer learning) and for identifying an individual afterwards (two strides) is compatible with practical applications. A second interesting finding is that CNNs can extract meaningful features from large gait datasets without preliminary feature engineering. These extracted features can be transferred favorably to assist in recognizing gaits from much smaller datasets. Modern CNNs are thus proving to be extremely effective for classifying gait signals, as they do for video-based gait recognition [19, 20].

The results of this study require confirmation in a larger sample of individuals walking in real-life environments. That said, COP analysis offers a promising alternative to video-based methods in niche biometric applications. However, further investigations are required to bring the COP method closer to a commercial application.

**Table 1**

Hyperparameter tuning. Bayesian optimization was used to explore the hyperparameter space in 300 trials. The optimal choices are shown in the last two columns

| Hyperparameters | Method | Values | Best results | |
|---|---|---|---|---|
| | | | CNN | sepCNN |
| CNN architecture | | | | |
|    Filter size in layers | Choice | A: 15, 13, 11, 11, [11, 11, 11], 3, 2 | C | C |
| | | B: 11, 9, 7, 7, [7, 7, 7], 3, 2 | | |
| | | C: 9, 7, 5, 5, [5, 5, 5], 3, 2 | | |
|    Number of filters in layers | Choice | A: 16, 16, 32, 64, [64,64, 64],64, 128, (128) | B | D |
| | | B: 32, 32, 64, 128, [128, 128, 128], 128, 256, (256) | | |
| | | C: 64, 64, 128, 256, [256, 256, 256], 256, 512, (512) | | |
| | | D: 128, 128, 256, 512, [512, 512, 512], 512, 1024, (1024) | | |
|    Number of intermediate blocks | Choice | 0, 1, 2, 3 | 2 | 1 |
|    Top-layer configuration | Choice | A: Global Average Pooling + Dense (Softmax) | A | A |
| | | B: Flatten + Dense + Dropout + Dense (Softmax) | | |
| Weight initialization | | | | |
| | Choice | A: Glorot (Xavier) normal initializer | A | A |
| | | B: He normal initializer | | |
| Activation | | | | |
| | Choice | ReLU | Swish | Swish |
| | | LeakyReLU | | |
| | | PReLU | | |
| | | Trainable Swish | | |
| Regularization | | | | |
|    L2 lambda | Log-uniform | $10^{-7}$ to $10^{-3}$ | $1.33 \cdot 10^{-5}$ | $1.01 \cdot 10^{-7}$ |
| Optimization | | | | |
|    Initial learning rate | Log-uniform | 0.0002 to 0.004 | 0.00068 | 0.00111 |

**Table 2**

Classification performance. Correct classification rate (accuracy) of the best standard convolutional neural network (CNN) and the best separable CNN (sepCNN) in ten trials.

| Trial | CNN | sepCNN |
|---:|---:|---:|
| 1 | 0.998 | 1.000 |
| 2 | 1.000 | 1.000 |
| 3 | 0.976 | 0.998 |
| 4 | 1.000 | 0.999 |
| 5 | 0.998 | 0.999 |
| 6 | 0.999 | 0.998 |
| 7 | 0.998 | 0.903 |
| 8 | 0.999 | 0.999 |
| 9 | 0.999 | 1.000 |
| 10 | 0.998 | 0.998 |
| Median | **99.84%** | **99.89%** |
| First quartile | 99.82% | 99.82% |
| Third quartile | 99.87% | 99.94% |

**Table 3**

Results of the verification experiments. The ability of the best CNN to differentiate between users and impostors was evaluated in 4 x 25 trials. The 36 subjects were randomly assigned to users / impostors for each trial, with four different proportions.

| Authorized users | Impostors | AUC (median) | AUC (1st quartile) | AUC (3rd quartile) | EER (median) | EER (1st quartile) | EER (3rd quartile) |
|---|---|---|---|---|---|---|---|
| 10 | 26 | 0.99997 | 0.99994 | 0.99999 | 0.27% | 0.23% | 0.31% |
| 15 | 21 | 0.99997 | 0.99994 | 0.99998 | 0.25% | 0.19% | 0.33% |
| 20 | 16 | 0.99995 | 0.99991 | 0.99996 | 0.33% | 0.25% | 0.41% |
| 25 | 11 | 0.99996 | 0.99995 | 0.99999 | 0.32% | 0.23% | 0.39% |
| Average | | **0.99997** | | | **0.29%** | | |

AUC: area under the (receiver operating characteristic) curve. EER: equal error rate.

**Table 4**

Summary of representative studies in footstep recognition.

| Study | Subjects | Steps | Footwear | Feature | Classifier | Performance |
|-------|----------|-------|----------|---------|------------|-------------|
| Jung et al. 2004 [17] | 11 | 440 | Barefoot | Foot shape + COP trajectory | HMM | FRR: 1.36%<br>FAR: 0.14% |
| Suutala et al. 2007 [18] | 11 | 440 | Shod | Vertical GRF profile | SVM, MLP | ACC: 95% |
| Moustadikis et al. 2008 [15] | 40 | 2,800 | Shod | 3D GRF profile | SVM | ACC: 98.2% |
| Pataky et al. 2011 [16] | 104 | 1,040 | Barefoot | Plantar pressure pattern | KNN | ACC: 99.6% |
| Derlatka 2013 [14] | 142 | 2,500 | Shod | 3D GRF profile | KNN | ACC: 98% |
| Connor 2015 [23] | 92 | 3,000 | Barefoot & Shod | Mixed | KNN | ACC: 99.8% (Barefoot)<br>ACC: 99.5% (Shod) |
| This study, identification | 30 | 90,000 | Shod | COP trajectory | CNN | ACC: 99.9% |
| This study, verification | 36 | 108,000 | Shod | COP trajectory | CNN | EER: 0.3% |

ACC: accuracy (correct classification rate). CNN: convolutional neural network. COP: center of pressure. EER: equal error rate. FAR: false acceptance rate. FRR: false rejection rate. GRF: ground reaction force. HMM: hidden Markov model. KNN: k-nearest neighbors. MLP: multilayer perceptron. SVM: support vector machine.

**Figure captions**

**Fig. 1**. *Center-of-pressure trajectory of walking. Five consecutive gait cycles recorded by the instrumented treadmill are shown. The raw 500 Hz signal was low-pass filtered at 30 Hz and down-sampled at 50 Hz. The X position corresponds to the movements perpendicular to the direction of progression. The Y position corresponds to the movements parallel to the direction of progression.*

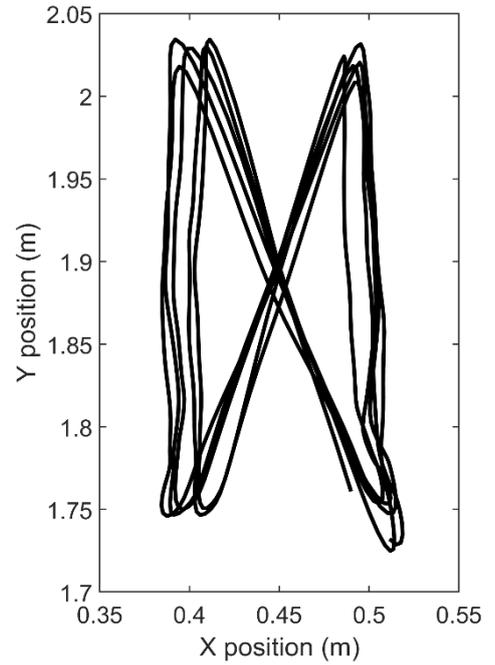

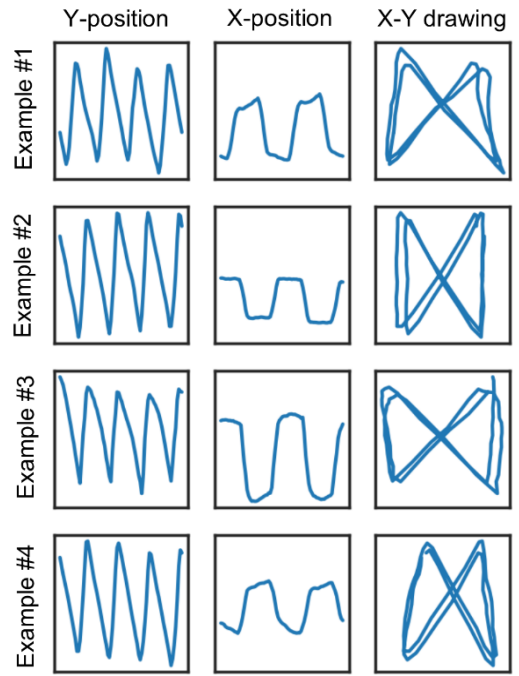

**Fig. 2.** *Examples of segments used to train the deep neural networks. Four examples from four distinct participants are shown. After time normalization at 40 samples per gait cycles, trajectory signals (Fig. 1) were sliced into 80-sample non-overlapping segments. These segments were fed into the first 1D convolutional layer as tensors of size (batch size ×80 ×2).*

**Fig. 3.** *Architecture of the deep convolutional neural network (CNN). Overall drawing that displays the general characteristics of the CNNs. The arrows show the residual shortcuts (ResNet). The number of intermediate blocks was adjusted during the hyperparameter tuning procedure. The weights of the trainable block were tuned in the transfer-learning analysis. For a detailed drawing of the final models see the supplementary online files.*

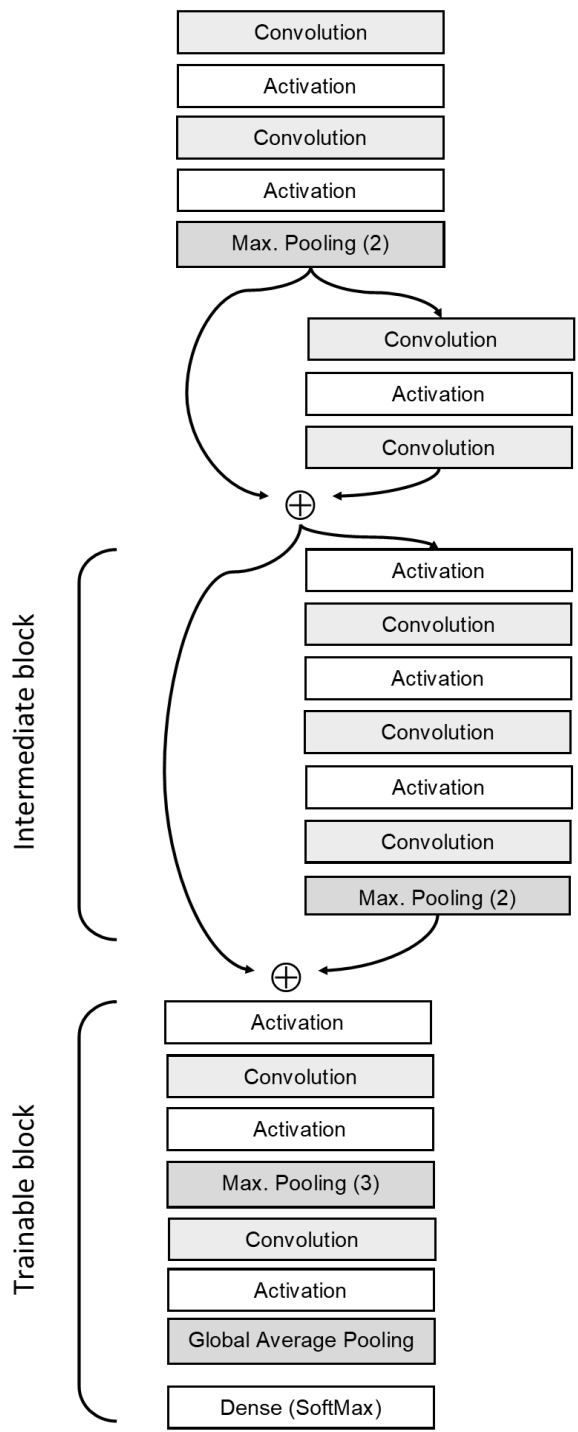

**Fig. 4.** *T-distributed stochastic neighbor embedding (t-sne) analysis of the last outputs of the CNNs used in transfer learning. The parameters of the best CNNs (Table 1) were frozen except for the last block of convolutional layers (Trainable block, Fig. 3). After training on the training set containing the gait data of six previously unseen individuals (4,050 segments), the fine-tuned CNNs were fed with the test set (450 segments). The flattened outputs of the last convolutional layers (labeled -1, -2 and -3) were analyzed through t-sne to highlight the separation of the features. Note that the -3 output corresponds to the output of the non-trainable part of the CNNs. Marker style and brightness correspond to the six individuals.*

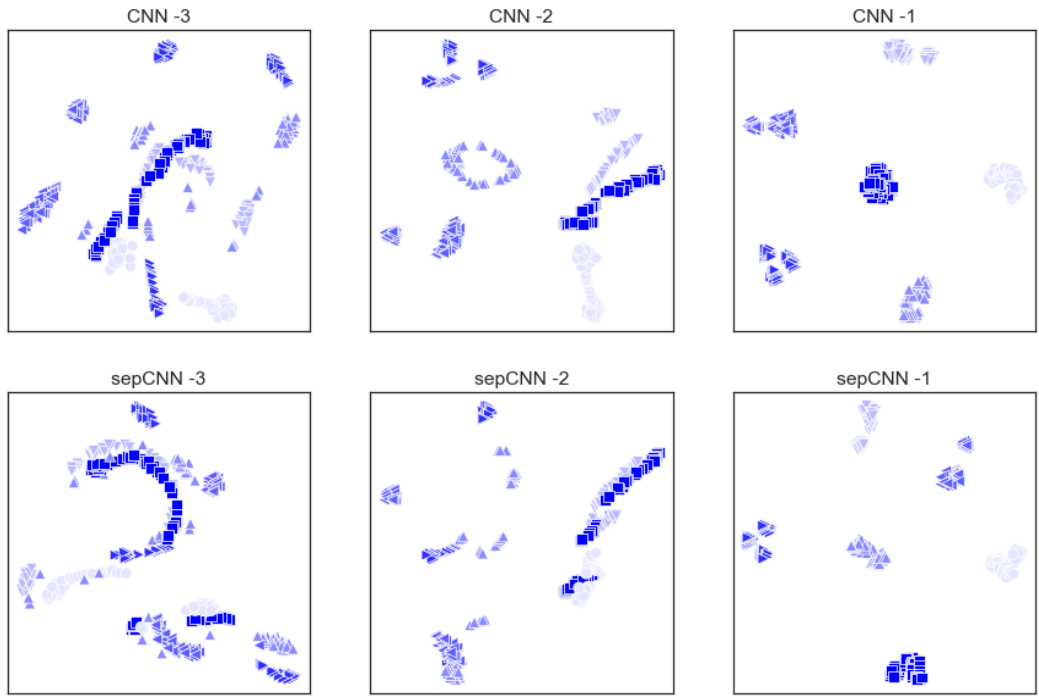

**Fig. 5.** *Transfer learning results. Six to sixty segments (1 to 10 per subject, x-axis) were drawn randomly from the training set and used to fine-tune CNNs. Overall classification accuracy (y-axis) was then computed on 60 segments drawn randomly from the test set. Fifty trials per segment number were finally repeated (total: 200 trials). Boxplots (quartiles and median) are shown. "+" indicates outliers. Note that the boxplots are collapsed because most of the trials reached 100% accuracy.*

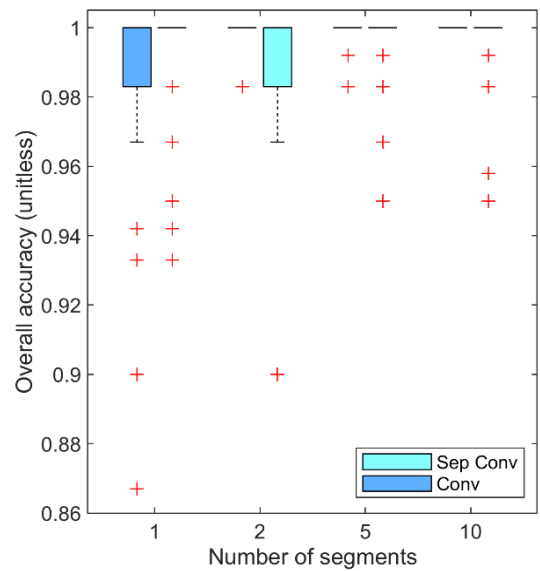